\documentclass[showpacs,preprint,aps]{revtex4}
\usepackage{graphicx}% Include figure files
\usepackage{dcolumn}% Align table columns on decimal point
\usepackage{bm}% bold math

\begin{document}
\begin{center}
{\bf Signature of the N=126 shell closure in dwell times of alpha-particle tunneling}
\\
N. G. Kelkar and M. Nowakowski\\
Departamento de Fisica, Universidad de los Andes,
Cra.1E No.18A-10, Bogot\'a, Colombia
\end{center}
\begin{abstract}
Characteristic quantities such as the penetration and preformation probabilities, 
assault frequency and tunneling times in the tunneling description of alpha 
decay of heavy nuclei are explored 
to reveal their sensitivity to neutron numbers in the vicinity of the magic neutron 
number $N$ = 126. Using realistic nuclear potentials, the sensitivity of these 
quantities to the parameters of the theoretical approach is also tested. 
An investigation of the region from $N=116$ to $N=132$ in Po nuclei reveals 
that the tunneling $\alpha$ particle spends the least amount of time with an 
$N=126$ magic daughter nucleus. The shell closure at $N=126$ seems to 
affect the behaviour of the dwell times of the tunneling alpha particles and this occurs 
through the influence of the $Q$-values involved. 
\end{abstract}
\maketitle
\section{Magic numbers and alpha decay of nuclei}
One of the most interesting findings of the early years of nuclear physics was the 
discovery of the existence of magic numbers which found an explanation based on the 
shell model of the nucleus. Even if a great number of magic nuclei have been 
identified over the years, 
their production and detailed study involves technical challenges and the subject as 
such continues to be a topic of current interest. 
The standardly recognized magic numbers with neutron or proton numbers of 2, 8, 20, 28, 
50, 82, 126, as such, originate from a spherical shell model 
but a deformed shell model also 
generates magic numbers with extra stability corresponding to the deformed structure. 
Indeed, based on empirical evidence, the authors in \cite{sureshK} while studying 
``superdeformed" nuclei 
identify $Z = 30, 38 - 41, 46, 58, 59, 62$ (and some others depending on the 
deformation) as magic numbers for nuclei with 
superdeformed shapes. In recent years, with the advent of radioactive 
beams, the experimental studies have been extended to the extremes of stability. 
These studies indicate that the shell structure established for nuclei near the 
$\beta$-stability line may change a lot for exotic nuclei. For instance, the neutron
numbers of 8, 20 and 28 are not magic in $^{12}$Be $(N=8)$, $^{32}$Mg $(N=20)$ and 
$^{42}$Si $(N=28)$ whereas new magic numbers such as $N=16, 24$ emerge in $^{24}$O
and $^{54}$Ca \cite{exoticnuclei}. In \cite{maitreyee}, a new shell closure at 
$N$ = 90 was found for neutron rich Sn isotopes \cite{maitreyee2}.

There exists yet another topic in nuclear physics which enjoys this kind of continued 
interest and this is the alpha decay of radioactive nuclei. Indeed, alpha 
decay was the very first application of the most exotic phenomenon of 
quantum mechanics, namely, tunneling as shown by Gamow \cite{gamow} and 
Condon and Gurney \cite{condon} in their pioneering works in 1928. 
We have come a long way since the discovery of magic
numbers and alpha-tunneling but the interest in the two phenomena taken together 
seems to be ever increasing \cite{paper1}. 
For example, in \cite{xuren1} it was found that even if 
empirical laws are usually sufficient to 
explain the alpha decay half-lives without a consideration of the alpha cluster 
preformation problem, close to the neutron shell closure of $N = 126$, one 
must take into account the shell model and other effects \cite{xuren1}. 

The present work is aimed at studying the behaviour of the characteristic quantities,  
appearing in the semiclassical approaches used to describe the alpha decay of 
heavy nuclei, as a function of the neutron number of the parent nuclei. The idea 
behind the investigation is to test if one or more of these quantities turn out to 
be good indicators of the existing magic numbers. In case they do, one could consider 
them to be tools for identifying possible shell closures.
In particular, the alpha decay 
of Polonium isotopes in the region around $N$ = 126 
is studied within a standard semiclassical approach involving 
the tunneling of a preformed alpha through the Coulomb barrier (for other 
approaches, see \cite{us1}). 
The tunneling probability, cluster preformation factor, assault frequency 
of the alpha at the barrier and the dwell time are calculated using 
a nuclear potential which is 
based on a double folding model with realistic nucleon-nucleon interactions
\cite{M3Y}. The potential is fitted to scattering data and
has also been tested in the $\alpha$ decay of several nuclei 
\cite{us1, us2}. In the next section we discuss the theoretical approach used 
for evaluating the decay width and the different tunneling time concepts used. 
An interpretation of the results is given in section 3. 

\section{Decay widths and tunneling times}
The general formula for the lifetime
of a nucleus decaying by $\alpha$-decay was obtained on the basis of a
Gamow-state formalism \cite{kadmen} in the seventies. Though such formalisms
\cite{nazar} are surely better than a semiclassical JWKB approach in general, 
for the purpose of the present investigations, it suffices to use the decay widths 
evaluated within the JWKB approximation \cite{width,froeman}. 
\subsection{Alpha-nucleus potential and the JWKB width} 
As shown in \cite{us2}, different semiclassical approaches lead to one 
and the same expression for the decay width, given by, 
\begin{equation} \label{width}
\Gamma(Q)=P_{\alpha}  \,\biggl [  
\frac{\hbar}{2\mu}\left[\int_{r_1}^{r_2}\frac{dr}{k(r)}\right]^{-1} \biggr ]\, 
P,
\end{equation} 
where $k(r)=\sqrt{2\mu(Q-V(r))}$, $Q$ (the $Q$-value in tunneling) 
is the amount of energy released in the decay 
and $P$ the tunneling probability in the JWKB approximation, i.e.,
\begin{equation} \label{prob}
P=\exp\biggl [ -2\int_{r_2}^{r_3}\kappa(r)dr \biggr ],
\end{equation} 
with, 
%\begin{equation} \label{kappa}
$\kappa(r)=\sqrt{2\mu(V(r)-Q)}$.
%\end{equation} 
$P_{\alpha}$ is the preformation probability of the $\alpha$ in the parent nucleus
and is often chosen as a free parameter to fit the theoretical half lives to the 
experimental ones, i.e. $P_{\alpha} = t_{1/2}^{theory}/t_{1/2}^{exp}$ where 
$ t_{1/2}^{theory} = \hbar \rm{ln} 2/ \Gamma $ (with $\Gamma$ evaluated as in 
Eq.(\ref{width}) but assuming $P_{\alpha} = 1$). 
The total potential $V(r)$ is a sum of the centrifugal (CF), 
nuclear (N)  and Coulomb (C) potentials \cite{us2}.  
The $\alpha$ is restricted within the classical turning points $r_1$, 
$r_2$ and $r_3$ defined by $V(r)$ and the Q values \cite{us2}. 
The factor in square brackets appearing before the penetration probability P 
in (\ref{width}) arises due to the 
normalization of the bound state wave function in the region from $r_1$ to $r_2$. 
The $\alpha$ is considered to tunnel through the potential 
\begin{equation} \label{pot}
V(r) = V_{C}(r) +  \lambda V_N(r) + \frac{\hbar (l+1/2)^2}{\mu r^2} \, , 
\end{equation}
where the strength $\lambda$ of the nuclear part $V_N$ is fixed by the 
Bohr Sommerfeld quantization condition \cite{us2}: 
\begin{equation} \label{Sommerfeld}
\int_{r_1}^{r_2} \,\, \sqrt{{2\mu\over \hbar^2 }
\,|V(r)\,-\,Q|}\,dr\,=\,(n\,+\,1/2)\,\pi.
\end{equation} 
Here, 
%\begin{equation} \label{n}
$n\, = \,(G\,-\,l)\,/2$, 
%\end{equation} 
is the number of nodes of the quasibound wave function of 
$\alpha$-nucleus relative motion and $r_1$ and $r_2$ which 
are solutions of $V(r) = Q$ are the classical turning points. 
$G$ is a parameter fitted to data which we will
discuss in more detail below.
$V_C$ in (\ref{pot}) is the Coulomb potential between the $\alpha$ and the daughter 
nucleus. The last term in $V(r)$ represents the Langer modified centrifugal 
barrier \cite{langer}. The modification from the standard 
$l(l+1) \rightarrow (l + 1/2)^2$ is required to ensure the correct behaviour of the 
JWKB scattered radial wave function near the origin as well as the validity of the 
connection formulas used. This modification leads to a potential which ``appears"
to have a centrifugal part even for $l=0$ as can be seen in the case of 
the $^{206}$Pb-$^4$He in Fig. 1. Other works on alpha decay using a similar approach 
can be found in \cite{alphanew,alphaold}. 
\begin{figure}[ht]
\includegraphics[width=8cm,height=7cm]{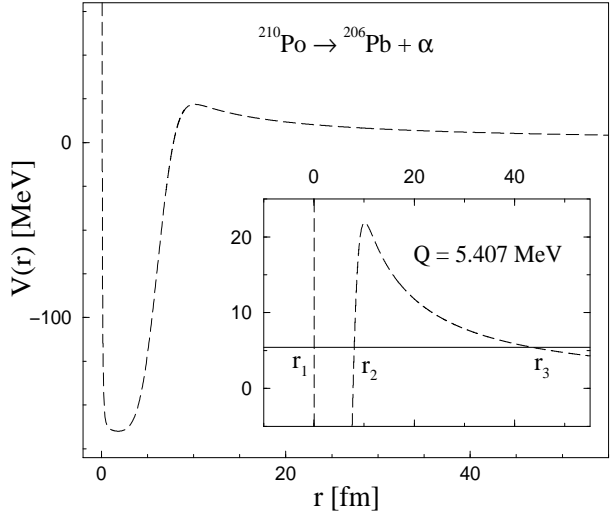}
\caption{\label{fig:eps1} Alpha-nucleus potential $V(r)$ as given in 
(\ref{pot}) for the $^{206}$Pb-$^4$He system resulting from the decay of $^{210}$Po. 
The inset displays the same potential with the classical turning points determined 
by the $Q$-value in the decay.}
\end{figure}

%The assault frequency (which in principle is a factor which arises due to the 
%normalization of the bound state wave function in the region from $r_1$ to $r_2$ 
%\cite{wkbrefs}) is given by, 
%\begin{equation} \label{nu}
%\nu=\frac{\hbar}{2\mu}\left[\int_{r_1}^{r_2}\frac{dr}{k(r)}\right]^{-1},
%\end{equation} 
%where $k(r)=\sqrt{2\mu(Q-V(r))}$. 

The folded nuclear part of the potential is given by a six dimensional integral
\begin{equation}\label{potnucl}
V_N({\bf r})\,= \,\int\,d{\bf r}_1\,d{\bf r}_2\,
\rho_{\alpha}({\bf r}_1) \, \rho_d({\bf r}_2)\,v({\bf r}_{12}\,=\,{\bf r}\,
+\,{\bf r}_2\,-\,{\bf r}_1,\,E)
\end{equation}
where $\rho_{\alpha}$ and $\rho_{d}$ are the densities of the alpha and the
daughter nucleus, respectively.
$v({\bf r}_{12},E)$ is the nucleon-nucleon 
interaction. $|\bf{r}_{12}|$ is the distance between a nucleon in the alpha 
and a nucleon in the daughter nucleus. 
We follow reference \cite{M3Y} which found that $v({\bf r}_{12},E)$ 
can be written as
\begin{eqnarray}\label{nnfree}
v({\bf r}_{12},E) \,&=&\,7999\,{exp(-4\,|{\bf r}_{12}|) 
\over 4\,|{\bf r}_{12}|} 
\, -\, 2134\, {exp(-2.5\,|{\bf r}_{12}|) \over 2.5\,|{\bf r}_{12}|}\,+\,
J_{00}\,\delta({\bf r}_{12})\\ \nonumber
J_{00}\,&=&\,-276\,(1\,-\,0.005\,E_{\alpha}/A_{\alpha})\,.
\end{eqnarray}
The respective densities entering equation (\ref{potnucl}) take the 
standard form  \cite{M3Y} 
\begin{equation}\label{alphadens}
\rho_{\alpha} (r)\,=\,0.4229\,exp(-0.7024\,r^2)
\end{equation}
and 
\begin{equation}\label{daughter}
\rho_d(r)\,=\, {\rho_0 \over 1\,+\,exp({r - c \over a})} \, .
\end{equation}
Here $\rho_0$ is obtained by normalizing $\rho_d(r)$ to the number of 
nucleons $A_d$ and the constants are given
as $c\,=\,1.07\,A_d^{1/3}$fm and $a\,=\,0.54$fm. 
Finally, denoting by $\rho^c_{\alpha}$ 
and $\rho^c_{d}$ the charge densities of the $\alpha$ and
daughter nucleus, the double folded the 
Coulomb potential is, 
\begin{equation}\label{potcol}
V_C(r)\,=\,\int\,d{\bf r}_1\,d{\bf r}_2\,
\rho^c_{\alpha}({\bf r}_1 \, \rho^c_{d}({\bf r}_2)\,{e^2 \over |{\bf r}_{12}|}
\,.
\end{equation}
The charge distributions are of similar form as the matter distributions above 
except for the normalization. The six dimensional integrals to evaluate the potentials 
can be made simpler by expressing the potential as an integral over the Fourier 
transforms of the densities and $v$. 
As a result one can write 
\begin{equation}
V_N(r) = {1 \over 2 \pi^2}\,  \int \, k^2 dk {\sin{(kr)} \over kr } \, v(k) \, 
\rho_{\alpha}(k) \, \rho_d(k) \, 
\end{equation}
for nuclei with $l = 0$ as considered in this work. $V_C(r)$ can be evaluated using 
similar methods.
The details of the procedure can be found in 
\cite{M3Y}. 

\subsection{Tunneling times and the assault frequency in alpha decay}
The concept of quantum tunneling times in connection with the half lives of 
radioactive nuclei was discussed in detail in \cite{us1}. Here we highlight 
the points relevant for the present work. 
Within a semiclassical picture one can define a ``period" $T$ as twice the time 
required for the $\alpha$ particle to traverse the region before 
the barrier, i.e. the distance between the turning points $r_1$ and $r_2$. The assault 
frequency is then the inverse of this quantity. 
Expressing the time interval, $\Delta t$, for the particle
traversing a distance, $\Delta r$ as,
\begin{equation}\label{2}
\Delta t \,=\, {\Delta r \over v(r)}\,=\,{\mu \, \Delta r \over \hbar\,k(r)}\,,
\end{equation}
the assault frequency $\nu$ can be written as the inverse of the time required
to traverse the distance back and forth between the turning points $r_1$ and
$r_2$ as \cite{froeman},
\begin{equation}\label{period}
\nu\,=\,T^{-1}\,=\,{\hbar \over 2\,\mu}\,\biggl[\,\int_{r_1}^{r_2}\, 
{dr \over 
\sqrt{{2\mu \over \hbar^2}\,(|V(r)\,-\,E|)}
}\,\biggr]^{-1}\,\,.
\end{equation}
This expression is however nothing but the ``normalization factor" which appeared 
in the square brackets in Eq. (\ref{width}). One can then rewrite Eq. (\ref{width}) as 
\begin{equation} \label{width2}
\Gamma(Q)=P_{\alpha} \nu P,
\end{equation} 
which is the often found form in literature. 

Interestingly, this quantity is directly proportional to the 
time spent by the $\alpha$-particle residing in the potential well. This 
time is also known as the dwell time $\tau_D$ \cite{butik} and 
is given by the number of particles in a given region of space divided by the 
incident flux. Thus, $\tau_D\,=\,{\int_{x_1}^{x_2}\,|\Psi(x)|^2\,dx / j}$  for 
a particle confined to the interval, $(x_1, x_2)$. 
The dwell time as such is considered to be a measure of the average time
spent by a particle in a given region of space. The concept was first
introduced by Smith \cite{smith} in the context of quantum collisions
and to derive a lifetime matrix for multichannel resonances. In the
one-dimensional case, it was first introduced by B\"uttiker \cite{butik}.
One can further define a ``transmission dwell time", $\tau_{D,T}$, 
corresponding to the time spent by those particles in a region (say, $(x_1, x_2)$, 
before the barrier) that managed to tunnel and 
get transmitted. As discussed in \cite{us1}, 
\begin{equation}
\tau_{D,T} = {\int_{x_1}^{x_2}\, |\Psi|^2 dx \over j_T}\, , 
\end{equation}
where, 
$j_T = \hbar \, k_0 |T|^2/ \mu$ with $k_0 = \sqrt{2 \mu E} / \hbar$ corresponding to 
the free particle energy $E$ and $|T|^2$ the transmission coefficient (which 
is the same as the penetration probability $P$ of the present work). 
In an investigation of the alpha decay half lives of heavy and super heavy nuclei, it 
was shown in \cite{us1}, that the JWKB decay width discussed in the previous sub-section 
is given by the inverse of the transmission dwell time in the region in front 
of the barrier:
\begin{equation} 
\Gamma = P_{\alpha} \, [\tau_{D,T}]^{-1} \, ,
\end{equation}
where $\tau_{D,T}$ is the transmission dwell time of the $\alpha$ in the 
region between $r_1$ and $r_2$. 

The half life of the decaying nucleus is thus given by
\begin{equation}\label{result2} 
\tau_{1/2} \, =\, {\hbar \, {\rm ln \,2} \over \Gamma} \, = 
\, {\hbar \, {\rm ln \,2} \over P_{\alpha} \, [\tau_{D,T}]^{-1}}  
\end{equation}
with $[\tau_{D,T}]^{-1} = \nu \, P$. Finally, we must mention that it can also be 
shown \cite{us1} that $\tau_D = 2 \tau_{\rm trav}$ where $\tau_{\rm trav}$ is the 
traversal 
time defined by B\"uttiker and Landauer \cite{butikland}. 
Therefore by calculating the half-lives we make a direct connection
to the quantum mechanical tunneling time concept. 
The behaviour of the alpha decay half lives displays a dependence on the magic numbers 
and hence through the present work we are attempting to relate the magic numbers to 
the quantum tunneling times in alpha decay. 

Before moving on to the next section, a brief discussion regarding the status of 
tunneling times is in order. 
A recent review \cite{bertulani} discusses resonant tunneling, 
tunneling of composite particles and how the coupling to intrinsic and external degrees 
of freedom can affect the tunneling probabilities. 
There exist extensive reviews 
\cite{hauge,nussen, bertsch} on the subject and one often 
finds contradictory remarks regarding the physical interpretation of some
of the times. Ref. \cite{nussen} for example considers the problem of tunneling time 
ill posed but that of an average dwell time well defined. Some of the controversies 
have arisen due to the Hartman effect \cite{harty} (the saturation of the phase time 
with increasing width of the barrier) which leads to interpretations based on 
superluminal propagation. The transmission dwell time as defined in the present work 
however does not lead to any such controversies \cite{goto} but can rather be related to 
lifetimes of nuclei as shown in \cite{us1}. The fact that the dwell time and not the 
phase time emerges as a useful concept was also shown in the 
context of eta-mesic nuclei in \cite{meprl}. In fact, an 
extraction of the dwell times as done in \cite{meapl} from 
the current-voltage characteristics in solid state tunnel junctions led to values 
very close to those measured from sophisticated experiments \cite{dwellexp}.
Given the fact that the dwell time seems to be emerging as the time concept with 
a physical meaning, it is timely to investigate in what other ways it manifests itself 
in tunneling processes. In this sense it is gratifying to find out that the
signature of shell closure can also be found in this concept.

\section{Discussion and Interpretation} 
Within the model described in the previous sections, we calculate the various
characteristic quantities in the tunneling of alpha and study their behaviour in the 
region of the neutron magic number $N = 126$ for Po isotopes. 

\subsection{Assault frequencies, $G$ value dependence and magic numbers} 
In Figure 2(d) we plot the assault frequency $\nu$ for Po isotopes 
as a function of the parent neutron numbers, 
using two different choices of the $G$ values appearing in the Bohr Sommerfeld 
condition (\ref{Sommerfeld}) on the potential. 
The lower curve represents the results using realistic potentials and the formulae
mentioned above but with the values of $G_<$, $G_>$, as in
\cite{paper1} where a toy model was introduced in order to study the behaviour 
of the assault frequencies as a function of the neutron number. 
In spite of the different nuclear input and the different definition of 
$\nu$ we reproduce the minimum at $N=126$ (note that the 
jump from $N = 126$ to $128$ in $\nu_M$ of the authors in 
\cite{paper1} is obvious since $\nu_M \propto G$ in their model and 
the authors choose to change $G$ from 20 to 24 for $N > 126$).  
The upper curve in Figure 2(d) here uses the $G$ value recommendations 
obtained from a fit \cite{Buck} to half lives 
of several nuclei. In \cite{Buck} the authors find that $G$ should be large (in the 
range 18 - 24) and that it should increase by two units while going from below the 
neutron magic number $N$ = 126 to above it. Thus, if $G = G_<$ 
corresponds to $82 < N \le 126$ and $G = G_>$ to $N > 126$, then 
$\Delta G = G_> 
- G_<$ should be always equal to 2. Hence we choose, 
$G=22$ for $N \le 126$, $G=24$ for $N > 126$. 
%and
%$G=23$ for $N=127$ (since $l=5$ in this case and we need an odd $G$). 
With this choice the assault frequency with realistic potentials 
changes and the minimum occurs now around $N=128 - 130$.
With such a sensitivity at hand $\nu$  might not be the
best indicator for a magic neutron number. 

\subsection{Cluster preformation probability}
In Figure 2(c), we also show the preformation probability, 
$P_{\alpha}=\ln 2/(\nu P \tau^{exp})$,  
based on the ratio between the theoretical, 
\begin{equation}\label{timeth}
\tau^{\rm theory} = \ln 2\, \, \tau_{D, T} = {\ln2 \over \nu P} \, ,
\end{equation}
and the experimental half lives. 
Recall here that $\tau_{D, T}$ is the transmission dwell time of the alpha in 
the region in front of the barrier. 
As in the case of $\nu$, we see that 
$P_{\alpha}$ is, in general, also sensitive to the
choice of $G$. However, as long as we restrict ourselves to even $N$  
the qualitative behaviour displaying a local minimum
at $N=126$ remains unchanged. 
We note that there is a clear model
dependence in $P_{\alpha}$ as far as the magnitude is concerned:
we differ from Refs \cite{paper1} by one order of magnitude. 
The order of magnitude of $P_{\alpha}$ here is similar to that in \cite{xuren1}.

\subsection{Dwell times in the vicinity of the $N=126$ shell closure}
In view of the sensitivity mentioned above,
it might be a good idea to look at the behaviour of
other quantities related to the $\alpha$-decay and
attempt a new interpretation. 
The curves for half lives in Figure 2(a)  
have a local maximum (minimum in the tunneling probability $P$) 
at $N=125$ and minimum (maximum in $P$) at $N=128$. 
To interpret this behaviour it is not enough to say
that $N=125$ is close enough to $N=126$ and therefore the 
maximum is intimately related to the magic number.
Indeed, one does not expect the magic nucleus to be the most
stable and therefore one would also not expect a maximum in lifetime
at the magic number. It is rather the steep slope starting
at $N=126$ to the next nucleus ($N=127$) (passing 
over orders of magnitude) which is a clear indicator
of a magic number since on account of the shell closure
one would expect the next nucleus to be much more unstable.

\begin{figure}[ht]
\includegraphics[width=12cm,height=10cm]{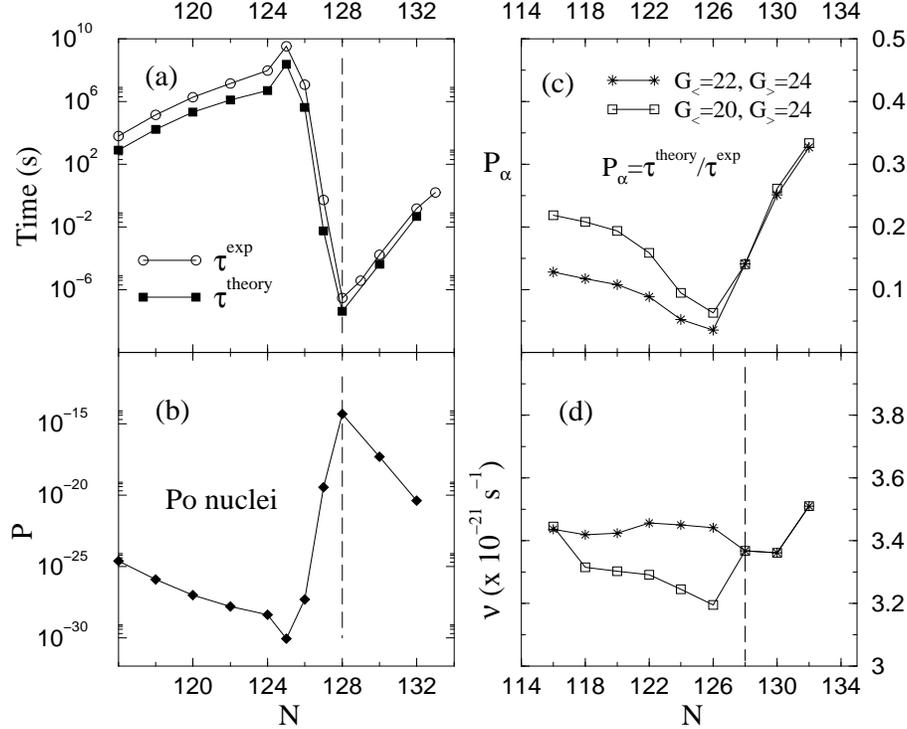}
\caption{\label{fig:eps2} Tunneling variables as a function of the parent 
neutron number. In (a) we see the experimental half lives and the theoretical ones 
with $\tau^{\rm theory} = \ln 2 \, \tau_{D, T} = \ln2 / \nu P$, (b) the penetration 
probability $P$ as in (\ref{prob}), (c) 
the preformation factor $P_{\alpha}$ and (d) the assault frequency $\nu$,  for $Po$
isotopes.}
\end{figure}
Thus the minimum in half lives 
occurring at $N=128$ has also to do with a magic number, but 
not of the parent nucleus. For $N=128$ of the parent nucleus
the daughter has the magic number $N=126$. 
The theoretical half-lives in Fig. 2(a), 
which are related to the transmission dwell time of the tunneling alpha, display a
similar behaviour. 

From the example considered above, it seems that the shell closure at $N=126$ 
influences the dwell times of the tunneling alphas. The answer as to why the shell 
closure should affect the dwell time is probably hidden in the dependence of these
times on the $Q$-value in the decay. Just as the binding energy of the last neutron,  
electric quadrupole moments and excitation energies from the ground to the first excited 
state can be correlated with shell closures, the $Q$-values in the alpha decays can 
also be 
considered as the messengers of the information regarding shell closures (as is evident 
in Fig. 3 discussed below). 
Thus it appears to us that an extension of the calculations in the present work but 
for other nuclei and in other regions of 
neutron numbers could possibly reveal similar effects of shell closures on the 
dwell times of the tunneling alpha particles. 

What seems so convincing for the neutron magic number, namely the fact that the 
parent nucleus ``prefers to decay into a magic daughter", is not a general feature 
when we probe into the proton magic numbers. 
In order to explore this point further, in Fig. 3, we plot the experimental half lives 
of nuclei as a function of the neutron and proton numbers in the vicinity of 
$N = 84$, $N =128$ (Fig 3(b)) and $Z = 84$ (Fig. 3(d)) 
which correspond to the parents of magic daughters. 
Though one observes dips in half lives at $N = 84$ and $N = 126$ no such structure 
is seen in the case of $Z = 84$. Even if a possible explanation could be that the Coulomb 
barrier height changes with changing $Z$, it is not obvious why a parent with 
$Z = 84$ would not decay rapidly to a magic daughter with $Z = 82$ (as seen in 
the left panel, Fig. 3(b), for magic $N = 82$).
Although it is beyond the scope of the present work, we think that the 
different behaviour of neutron and proton magic numbers studied from the 
point of view of tunneling times calls for an explanation. 
\begin{figure}[ht]
\includegraphics[width=12cm,height=10cm]{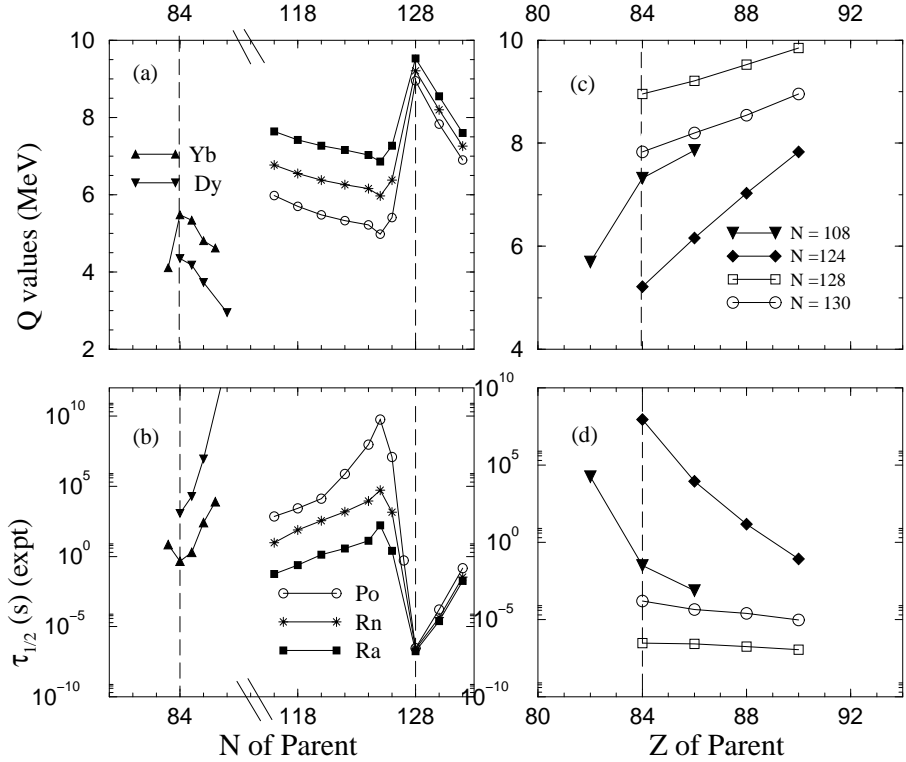}
\caption{\label{fig:eps3} Q values and experimental half lives \cite{tuli} 
of nuclei as a function 
of the neutron and proton numbers of the parent nuclei.}
\end{figure}

\section{Summary}
The alpha decay of heavy nuclei is commonly treated within a model where the 
decaying parent nucleus is made up of a cluster of an alpha or $^4$He nucleus and
and the daughter in the decay. The decay is possible due to the tunneling of the alpha 
through the Coulomb barrier produced due to its interaction with the daughter. The 
present work studies the characteristic quantities in such a tunneling process and 
in particular their behaviour in the region around the neutron magic number $N = 126$. 
Calculations using realistic nuclear potentials are presented for isotopes of 
the Polonium nucleus in the region of neutron numbers $N = 116$ to $N = 132$. 
The findings of this work can be summarized as follows:
\begin{itemize}
\item [(i)] The amount of time spent by an alpha in front of the barrier before tunneling 
(the transmission dwell time discussed in section 2, $\tau^{\rm theory}$ in Fig. 2) 
reaches a minimum at $N = 128$ 
of the parent nucleus in the region from $N = 116$ to $N= 132$ studied in this work. 
$N = 128$ of the parent however corresponds to $N= 126$ of the daughter, implying that 
the alpha spends the least amount of time with the magic daughter. 
\item[(ii)] The frequency of assaults $\nu$ of the alpha at the barrier is  
found to be sensitive to the 
parameter $G$ used in the semiclassical JWKB approach. Though for a certain set 
of $G_<$, $G_>$ (defined in section 3) the minimum in $\nu$ occurs at the magic
number $N = 126$ of the parent, this can change with a small change in the values 
of $G_<$, $G_>$. 
\item 
[(iii)] The assault frequency is shown to be related to the ``traversal time" 
concept defined in \cite{butikland}, whereas the half life of the decaying nucleus 
is related to the ``transmission dwell time" defined in \cite{us1}. 
From the present calculations, it seems that the transmission dwell time is 
a clearer indicator of the magic number $N = 126$ rather than the assault frequency 
or traversal time which are sensitive to the parameters in the JWKB approach.
\end{itemize}
Though intuitively, one would expect a similar behaviour of the 
tunneling characteristics for other heavy nuclei,  
it would be interesting to investigate other nuclei as well as 
other regions of magic numbers within the approach of the present work in order 
to confirm the conclusions of this work.  

\begin{acknowledgments}
We thank the  administrative  department  of  science,  technology  and  innovation of Colombia (COLCIENCIAS) for  the financial support provided.
\end{acknowledgments}


\begin{thebibliography}{99}
\bibitem{sureshK}
Neha Sharma, H. M. Mittal, Suresh Kumar and A. K. Jain, Phys. Rev. C {\bf 87}, 
024322 (2013).
\bibitem{exoticnuclei}
B. Bastin {\it et al.}, Phys. Rev. Lett. {\bf 99}, 022503 (2007); 
Z. Y. Xu {\it et al}., Phys. Rev. Lett. {\bf 113}, 032505 (2014); 
H. Iwasaki {\it et al}., Phys. Lett. B {\bf 481}, 7 (2000).
\bibitem{maitreyee}
S. Sarkar and M. Saha Sarkar, Phys. Rev. C {\bf 81}, 064328 (2010).
\bibitem{maitreyee2}
S. Sarkar and M. Saha Sarkar, Phys. Rev. C {\bf 78}, 024308 (2008); S. Sarkar and
M. Saha Sarkar, Phys. Rev. C {\bf 64}, 014312 (2001). 
\bibitem{gamow}
G. Gamow, Z. Phys. {\bf 51}, 204 (1928).
\bibitem{condon}
R. W. Gurney and E. U. Condon, Phys. Rev. {\bf 33}, 127 (1929);
{\it ibid}, Nature {\bf 122}, 439 (1928).
\bibitem{paper1} H. F. Zhang, G. Royer and J. Q. Li, Phys. Rev. C {\bf 84}, 
027303 (2011); H. F. Zhang and G. Royer, Phys. Rev. C {\bf 77}, 054318 (2008). 
\bibitem{xuren1}
Chang Xu and Zhongzhou Ren, Phys. Rev. C {\bf 76}, 027303 (2007); 
D. Ni and Z. Ren, Ann. Phys. {\bf 358}, 108 (2015).
\bibitem{us1} 
N. G. Kelkar, H. M. Casta\~neda and M. Nowakowski, Europhys. Lett. {\bf 85}, 
20006 (2009).
\bibitem{M3Y}
G. R. Satchler and W. G. Love, Phys. Rep. {\bf 55}, 183 (1979);
A. M. Kobos, B. A. Brown, P. E. Hodgson, 
G. R. Satchler and A. Budzanowski, Nucl. Phys. {\bf A384}, 65
(1982).
\bibitem{us2} 
N. G. Kelkar and H. M. Casta\~neda, Phys. Rev. C {\bf 76}, 064605 (2007). 
\bibitem{kadmen}
S. G. Kadmensky, V. E. Kalechitz and A. A. Martynov,
Phys. At. Nucl. {\bf 16}, 717 (1972);
S. G. Kadmensky and V. I. Furman, Part. Nucl. {\bf 6}, 469 (1975);
S. G. Kadmensky and V. I. Furman, {\it Alpha Decay and Related Nuclear
Reactions}, Moscow, Energoatomizdat (1985).
\bibitem{nazar}
D. S. Delion, R. J. Liotta and R. Wyss, Phys. Rep. {\bf 424}, 113 (2006);
S. A. Gurvitz, P. B. Semmes, W. Nazarewicz and T. Vertse,
Phys. Rev. {\bf A69}, 042705 (2004).
\bibitem{width}
S. A. Gurvitz and G. Kalbermann, Phys. Rev. Lett. {\bf 59}, 262 (1987); 
S. A. Gurvitz, Phys. Rev. A {\bf 38}, 1747 (1988).
\bibitem{froeman}
G. Drukarev, N. Fr\"oman and P. O. Fr\"oman, J. Phys. {\bf A 12}, 171 (1979).
\bibitem{langer}
R. E. Langer, Phys. Rev. {\bf 51}, 669 (1937).
\bibitem{alphanew}
M. Ismail, A. Y. Ellithi, A. Adel and A. R. Abdulghany, Nucl. Phys. A {\bf 947}, 
64 (2016); 
Dongdong Ni and Zhongzhou Ren, Phys. Rev. C {\bf 93}, 054318 (2016). 
\bibitem{alphaold}
C. Xu and Z. Ren, Nucl. Phys. A {\bf 760}, 303 (2005); Dongdong Ni and Zhouzhou Ren, 
Nucl. Phys. A {\bf 825}, 145 (2009).
\bibitem{butik}
M. B\"uttiker, Phys. Rev. {\bf B 27}, 6178 (1983).
\bibitem{smith}
F. T. Smith, Phys. Rev. {\bf 118}, 349-356 (1960).
\bibitem{butikland}
M. B\"uttiker and R. Landauer, Phys. Rev. Lett. {\bf 49}, 1739 (1982).
\bibitem{bertulani}
C. A. Bertulani, Few Body Syst. {\bf 56}, 727 (2015); C. A. Bertulani, V. V. Flambaum 
and V. G. Zelevinsky, J. Phys. G {\bf 34}, 2289 (2007).
\bibitem{hauge}
E. H. Hauge and J. A. St\o vneng, Rev. Mod. Phys. {\bf 61}, 917 (1989);
J. G. Muga, R. Sala-Mayato and I. L. Egusquiza, {\it Time in
Quantum Mechanics}, New York: Springer (2002); R. Landauer and Th. Martin,
Rev. Mod. Phys. {\bf 66}, 217 (1994);
V. S. Olkhovsky and E. Recami, Phys. Rep. {\bf 214}, 339 (1992);
H. Winful, Phys. Rep. {\bf 436}, 1 (2006).
\bibitem{nussen}
C. A. A. de Carvalho and
H. M. Nussenzveig, Phys. Rep. {\bf 364}, 83 (2002). 
\bibitem{bertsch}
P. Arve, G. F. Bertsch, J. W. Negele and G. Puddu, Phys. Rev. C {\bf 36}, 2018 (1987). 
\bibitem{harty}
T. E. Hartman, J. Appl. Phys. {\bf 33},
3427 (1962).
\bibitem{goto}
M. Goto, H. Iwamoto, V. M. de Aquino, V. C. Aguilera-Navarro and
D. H. Kobe, J. Phys. {\bf A 37}, 3599 (2004).
\bibitem{meprl}
N. G. Kelkar, Phys. Rev. Lett. {\bf 99}, 210403 (2007). 
\bibitem{meapl}
E. J. Pati\~no and N. G. Kelkar, Apl. Phys. Lett. {\bf 107}, 253502 (2015).
\bibitem{dwellexp}
D. Shafir, H. Soifer, B. D. Bruner, M. Dagan, Y. Mairesse, S. Patchkovskii, M. Y. Ivanov, O. Smirnova, N. Dudovich, Nature {\bf 485}, 343 (2012); P. Eckle, A. N. Pfeiffer, C. Cirelli, A. Staudte, R. D\"{o}rner, H. G. Muller, M. B\"{u}ttiker, U. Keller, Science {\bf 322}, 1525 (2008).
\bibitem{Buck} B. Buck and A. C. Merchant, Phys. Rev. C {\bf 45}, 2247 (1992). 
\bibitem{tuli} Jagdish K. Tuli, Nuclear Wallet Cards, 8$^{th}$ edition, NNDC, 
Brookhaven National Laboratory, 2011 (see also: www.nndc.bnl.gov/wallet). 
\end{thebibliography}
\end{document}